\documentclass[options,onecolumn]{mn2e}
\usepackage{psfig}
\newif\ifAMStwofonts

\newcommand{\mt}[1]{\mbox{$\mathbfss{#1}$}}
\newcommand{\VEV}[1]{\langle#1\rangle}

\title[Cosmological parameters and the \emph{WMAP} data revisited]{Cosmological parameters and the \emph{WMAP} data revisited}
  
\author[F. K. Hansen, A. Balbi, A. J. Banday and K. M. G\'orski]
  {{F. K. Hansen$^1$, \thanks{E-mail: frodekh@roma2.infn.it}}, {A. Balbi$^{1,2}$, \thanks{E-mail: Amedeo.Balbi@roma2.infn.it}}, {A. J. Banday$^3$, \thanks{E-mail: banday@MPA-Garching.MPG.DE}}, {K. M. G\'orski$^{4}$, \thanks{E-mail: Krzysztof.M.Gorski@jpl.nasa.gov}}\\
$^1$ Dipartimento di Fisica, Universit\`a di Roma `Tor Vergata', Via della Ricerca Scientifica 1, I-00133 Roma, Italy\\
$^2$ INFN, Sezione di Roma II, Via della Ricerca Scientifica 1, 00133 Roma Italy\\
$^3$ Max Planck Institut f\"ur Astrophysik, Karl-Scharzschild-Strasse 1, Postfach 1317, D-85741 Garching bei M\"unchen, Germany\\
$^4$ JPL, M/S 169/327, 4800 Oak Grove Drive, Pasadena CA 91109 \\}

\pagerange{\pageref{firstpage}--\pageref{lastpage}}
\pubyear{2002}

\begin{document}

\label{firstpage}

\maketitle

\begin{abstract}
Several recent papers have made claims about the detection of an
asymmetric distribution of large scale power in the cosmic microwave
background anisotropy as measured by the \emph{WMAP} satellite. 
In this paper, we investigate how 
the estimates of particular cosmological parameters vary
when inferred from power spectra computed separately on
the northern and southern hemispheres in
three frames of reference: the galactic, the ecliptic and the 
specific frame of reference which maximises the asymmetry 
between the power spectra from the corresponding hemispheres.
Such a study is intended to quantify the consistency of the
observed spectral variations in the context of 
the inflationary-scenario-inspired models with which the data 
are commonly compared.
We focus our investigation on the three specific parameters to which
the analysis is most sensitive -- 
the spectral index $n$, the amplitude of fluctuations
$A$ and the optical depth $\tau$ --
and find interesting variations in their estimates as determined 
from  different hemispheres.  
When using a gaussian prior on the spectral index $n$ centred at $n=1$ with a flat prior on
the optical depth $\tau$, the preferred value for the optical depth
(derived in the reference frame of maximum asymmetry) 
in the north is $\tau=0$ with upper limit $\tau<0.08$, whereas in the
south we find $\tau=0.24^{+0.06}_{-0.07}$ ($68\%$ confidence level).
The latter result is inconsistent
with $\tau=0$ at the $2\sigma$ level. The estimated optical depth of
$\tau=0.17$ on the (nearly) full sky found by the \emph{WMAP} collaboration and
confirmed independently here, could thus in large part originate
in structure associated with the southern
hemisphere. Furthermore, putting a prior on $\tau$, we find values of
the spectral index which are inconsistent between the opposite
hemispheres. The exact values depend on the prior on $\tau$. 
Our conclusions remain unaltered even when,
on the basis of putative residual foreground contamination,
the multipole range $\ell=2-7$ is excluded from the analysis.
While our results should not be considered more than suggestive,
the significance of the parameter differences in the two hemispheres
being typically of order $2 \sigma$, if they are confirmed with
the higher sensitivity \emph{WMAP} 2-year data then it may be
necessary to question the assumption of cosmological isotropy
and the conceptual framework within which studies of the CMB
anisotropy are made.
\end{abstract}


\section{introduction}

Several recent papers have determined from an analysis
of the data from the \emph{WMAP} satellite \cite{WMAPref}
that the fluctuations in the cosmic microwave
background (CMB) are asymmetrically distributed on the sky
\cite{park,eriksen1,vielva,eriksen2,curvat,copi,p3,wandelt,cruz,land}. In
particular, it was shown \cite{eriksen1,p3} that the power spectra as
computed in the northern and southern hemispheres (in both the galactic and
ecliptic frames of reference) differ significantly on large angular
scales. Specifically, for multipoles between 5 and 40,
the southern hemisphere spectrum showed evidence of excess power
whereas the northern spectrum was remarkably lacking therein,
with respect to the canonical full sky estimate
\footnote{This was considered to be the best fit 
power-law spectrum, as determined from the \emph{WMAP} 
analysis using the Kp2 sky cut including the point source mask.
Whenever we refer to a `full-sky' analysis this coverage is implied,
and the nomenclature is simply for convenience}. 
The asymmetry in the spectrum over these multipoles was found to be significant at the
$2-3\sigma$ level compared to Gaussian realizations of the best fit
\emph{WMAP} power spectrum. Moreover, the axis of maximum asymmetry
was found to have its north pole at $(80^\circ,57^\circ)$ (galactic
\emph{co}-latitude and longitude). This
direction was subsequently confirmed by \cite{curvat,land}
using tests of non-Gaussianity. A similar (albeit at lower statistical
significance) asymmetric distribution of
power was also found using the data of the \emph{COBE}-DMR \cite{COBEref}
experiment. 
There have also been detections of non-Gaussianity in
the \emph{WMAP} data \cite{nas1,nas2,nas3,park,vielva,curvat,wandelt,land,eriksen2,cruz}.
Whilst some of these have been claimed to be
consistent with residual foreground signals \cite{nas1,nas2,nas3}, it
remains uncertain whether this non-Gaussianity is related to the power spectrum
asymmetry. Indeed, although indications of a possible foreground contamination of the
quadrupole and octupole have been found
(i.e. \cite{p3,schwarz,eriksen3,slosar}), the asymmetry at higher
multipoles is harder to explain in terms of incorrectly subtracted
foregrounds since it was found to be present at the same amplitude
in all the \emph{WMAP} frequencies and does not seem to decrease
significantly with an increasing galactic cut
\cite{eriksen1,p3}. 

In this paper, the aim is not to obtain high-precision estimates of
cosmological parameters in themselves, but rather to 
assess which parameters are more likely to be affected by the measured
asymmetry and then to investigate and quantify the
differences of the parameters between the north and the south.
We will restrict the class of models for which we determine these
parameterisations to the basic (inflationary-scenario-inspired)
cosmological model as considered by the \emph{WMAP} team
\cite{spergel} -- a flat $\Lambda$CDM model with power-law
adiabatic primordial perturbations. The parameters of this model are
the amplitude $A$ and the spectral index $n$ of the primordial power
spectrum of scalar perturbations (modelled as $Ak^n$), the optical
depth $\tau$, the physical density of baryons and cold dark matter,
$\Omega_b h^2$ and $\Omega_{cdm} h^2$ (where $h$ is the Hubble
constant in units of 100 km/s/Mpc), and the cosmological constant
$\Omega_\Lambda$. Note that when fixing $\Omega=1$, the Hubble
constant $h$ is set when the parameters $\Omega_b h^2$,
$\Omega_{cdm} h^2$ and $\Omega_\Lambda$ are given.  We determine
estimates of these parameters in the northern and southern hemisphere
in three reference frames: galactic, ecliptic and that frame which maximises
the asymmetry in the power spectra computed in the corresponding
hemispheres.

\section{method}

In \cite{p3}, the power spectrum was estimated on differently
oriented hemispheres using the Gabor transform formalism introduced by
\cite{hansen1,hansen2} and extended in \cite{p3}. A brief
outline of the method is provided here.

A Gaussian likelihood ansatz is adopted with the pseudo power
spectrum as input. The pseudo power spectrum $\tilde C_\ell$ is given
by
\begin{equation}
\tilde C_\ell=\sum_{m=-\ell}^{\ell}\frac{\tilde a_{\ell m} \tilde a^*_{\ell m}}{2\ell+1},
\end{equation}
where
\begin{equation}
\tilde a_{\ell m}=\int T({\hat n}) G({\hat n}) d{\hat n}
\end{equation}
are the spherical harmonic coefficients on the observed sky. Here,
${\hat n}$ is the position on the sky, $T({\hat n})$ is the CMB
temperature fluctuation amplitude at that position and $G({\hat n})$ is the mask.
The latter quantity takes zero value in those parts of the sky which are either unobserved or 
have been excluded from the analysis, and is otherwise 
one. The likelihood can now be written as
\begin{equation}
\label{eq:multilik}
\mathcal{L}=\frac{\mathrm{e}^{-\frac{1}{2}{\mathbf{ d}}^\mathrm{T} {\mt{M}}^{-1}{\mathbf{d}}}}{\sqrt{2\pi \det{\mt{M}}}},
\end{equation}
where the elements of the data vector are
\begin{equation}
d_i=\tilde{C}_{\ell_i}-\VEV{\tilde{C}^S_{\ell_i}}-\VEV{\tilde{C}^N_{\ell_i}}.
\end{equation}
Here, the first term is the observed pseudo power spectrum including
noise and the second term is the expected signal pseudo spectrum as a
function of the underlying full sky power spectrum $C_\ell$. The last
term contains the expected noise pseudo spectrum for the observed
part of the sky. Similarly, the matrix expressing the correlation
between the pseudo power spectrum coefficients can be written as
\begin{equation} 
M_{ij}=<\tilde C_{\ell_i}\tilde C_{\ell_j}>-<\tilde C_{\ell_i}><\tilde
 C_{\ell_j}>=M_{ij}^\mathrm{S}+M_{ij}^\mathrm{N}+M_{ij}^\mathrm{X},
\end{equation}
where there is one contribution from the signal (S), one from the
noise (N) and one from signal-noise mixing (X). 
The exact form of these matrices,
their dependence on the full sky power spectrum $C_\ell$ and
details of their computation are discussed in
\cite{hansen1,hansen2,p3}. In these papers, the power spectrum was
estimated by maximising this likelihood with respect to $C_\ell$. As
the likelihood is not Gaussian with respect to the power spectrum,
there is a small bias introduced in the estimate of the $C_\ell$
\cite{bjk} at the smallest multipoles, for which 
a correction was found in \cite{p3}. In this paper, we maximise the likelihood
with respect to the cosmological parameters. We have tested this
method on simulated maps with known cosmological parameters and have
demonstrated that the procedure gives an unbiased estimate of the
input cosmological parameters.

\section{Preparing the data}

\label{sect:data}

In this paper, we analyse the publicly available \emph{WMAP} 
data\footnote{These can be obtained at the LAMBDA website:
\emph{http://lambda.gsfc.nasa.gov/}} after correction for Galactic foregrounds.
We consider data from the V and W frequency bands only
since these are less likely to be contaminated by residual foregrounds. 
The maps from the different channels are coadded using inverse
noise-weighting \cite{hinshaw}:
\[
T=\frac{\sum_cT_c/\sigma_c^2}{\sum_c1/\sigma_c^2},
\]
where $c$ runs over the frequency channels V1,V2,W1,W2,W3,W4. 
All analysis is performed using the \emph{WMAP} Kp2 sky cut, including the point source mask.

\section{the results}

Since we are interested in estimating the relative difference in parameters
between the hemispheres rather than an exact evaluation of their
values to high
precision, the multipoles used in the analysis are limited to the
ranges which are important for characterising the difference between
the northern and the southern spectra. The asymmetry manifests itself 
as both a difference in shape over the first 40 multipoles and as a difference in
amplitude between the low-order multipoles and those at higher $\ell$.
Therefore, multipoles in the range $\ell=2-60$ are selected 
to account for the shape of the spectrum at large
angular scales, together with a set of multipoles centred around the
first two peaks ($\ell=200-240$ and $\ell=520-560$) to constrain the
shape of the spectrum at smaller scales.
As noted in the introduction, we are generally interested in the parameters $A$, $n$, $\tau$,
$\Omega_b h^2$, $\Omega_\Lambda$ and $\Omega_{cdm}h^2$, but have found that
the asymmetry does not depend strongly on the exact value of
the latter two values.  
This should not be surprising, since
when $\Omega$ is fixed, the parameters $\Omega_\Lambda$ and
$\Omega_{cdm}h^2$ influence the position of the acoustic peaks in the
power spectrum, which does not differ significantly between the
different hemispheres \cite{p3}. Consequently, it was decided for simplicity to fix
these parameters to the best fit values found by the \emph{WMAP} team
($\Omega_\Lambda=0.74$ and $\Omega_{cdm} h^2=0.11$). These values
are supported by independent astronomical observations that
are consistent with the \emph{WMAP} findings (for example, large scale structure matter distribution
\cite{verde}, HST Key Project, \cite{freedman}, type Ia supernovae
\cite{riess,perlmutter}) and that are not likely to be affected by the
asymmetry.
In addition, we have explicitly checked that the results in
this paper are robust with respect to changes in these parameters.

\begin{figure}
\begin{center}
\leavevmode
\psfig {file=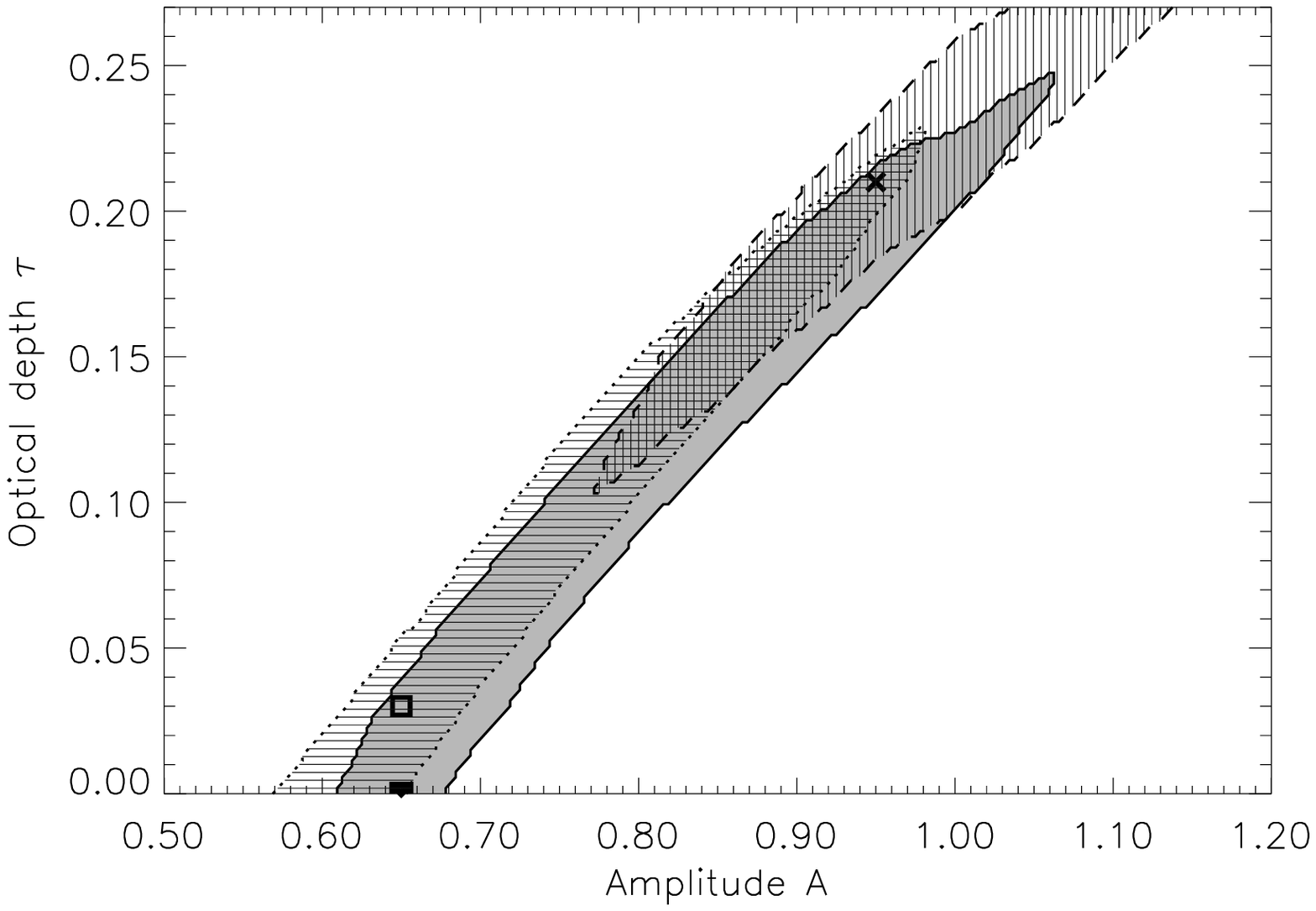,width=7cm,height=7cm}
\psfig {file=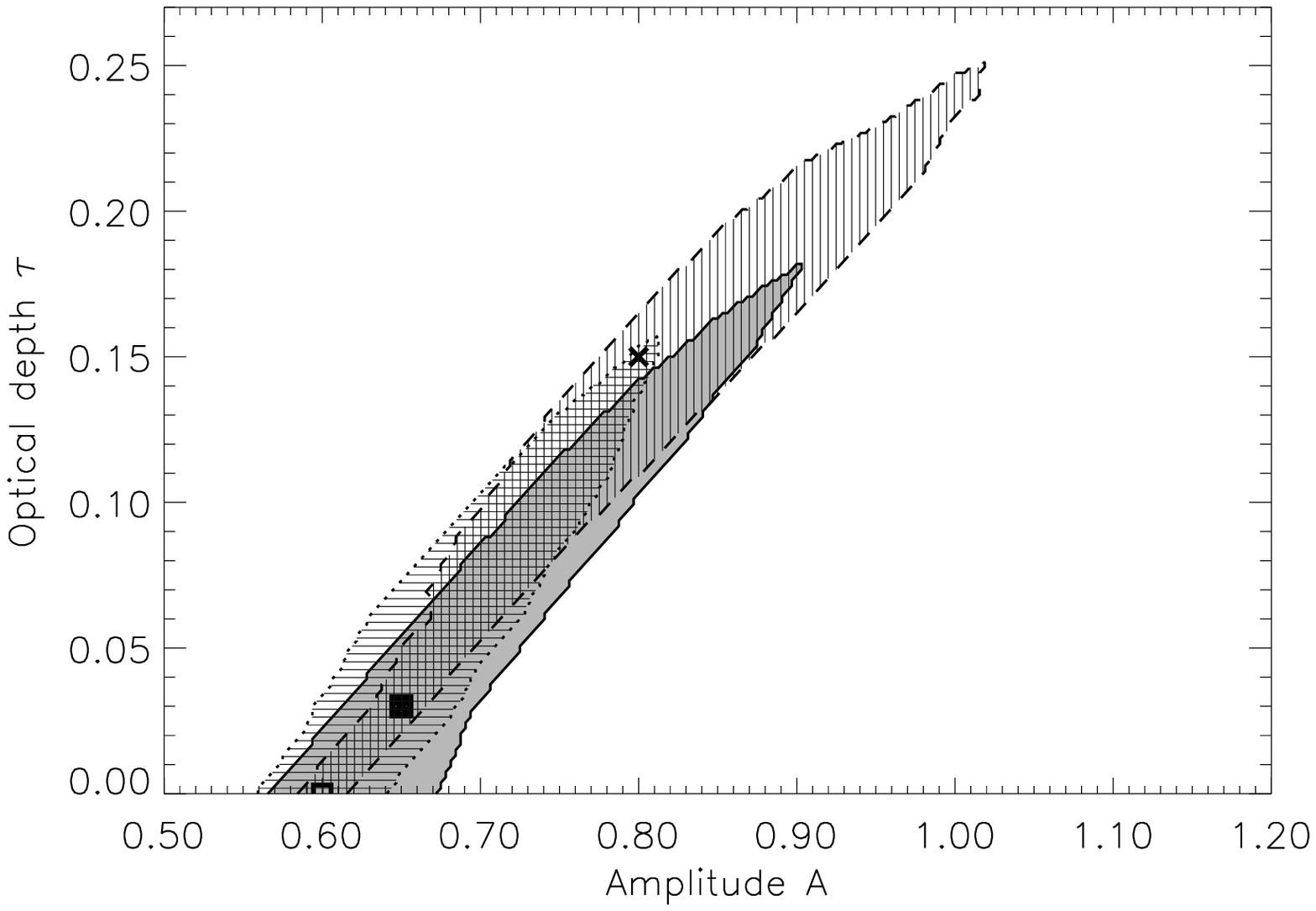,width=7cm,height=7cm}
\psfig {file=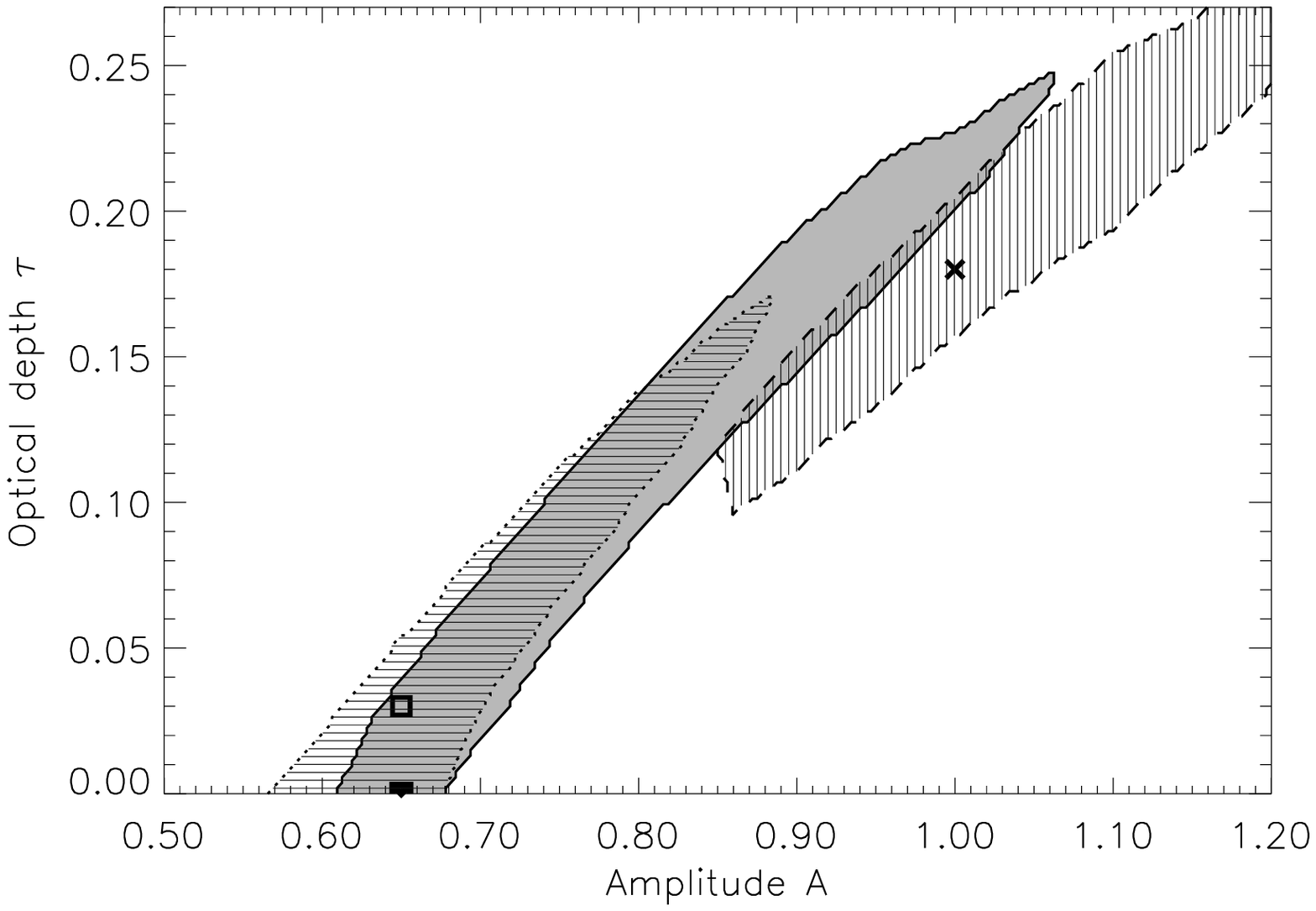,width=7cm,height=7cm}
\psfig {file=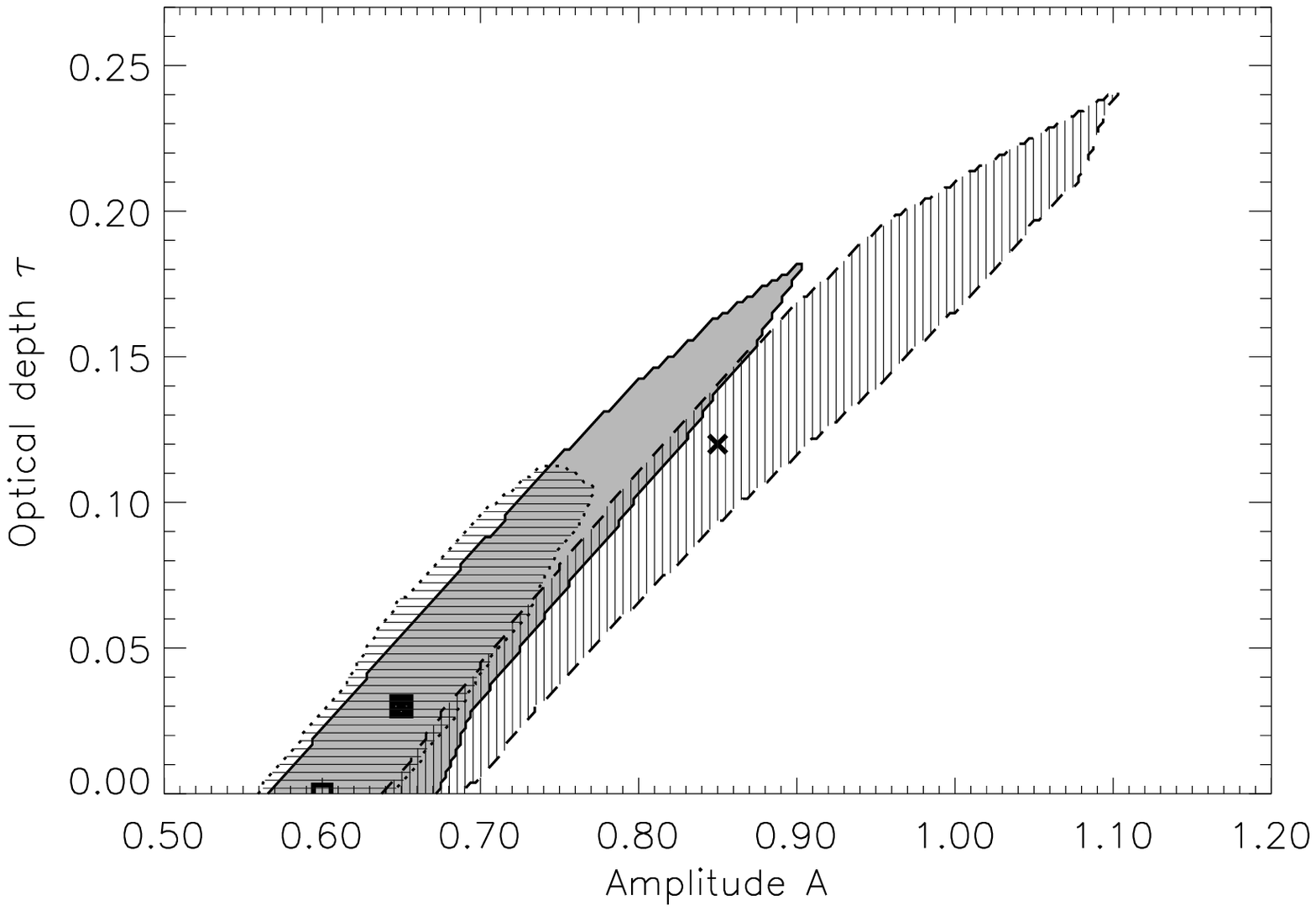,width=7cm,height=7cm}
\psfig {file=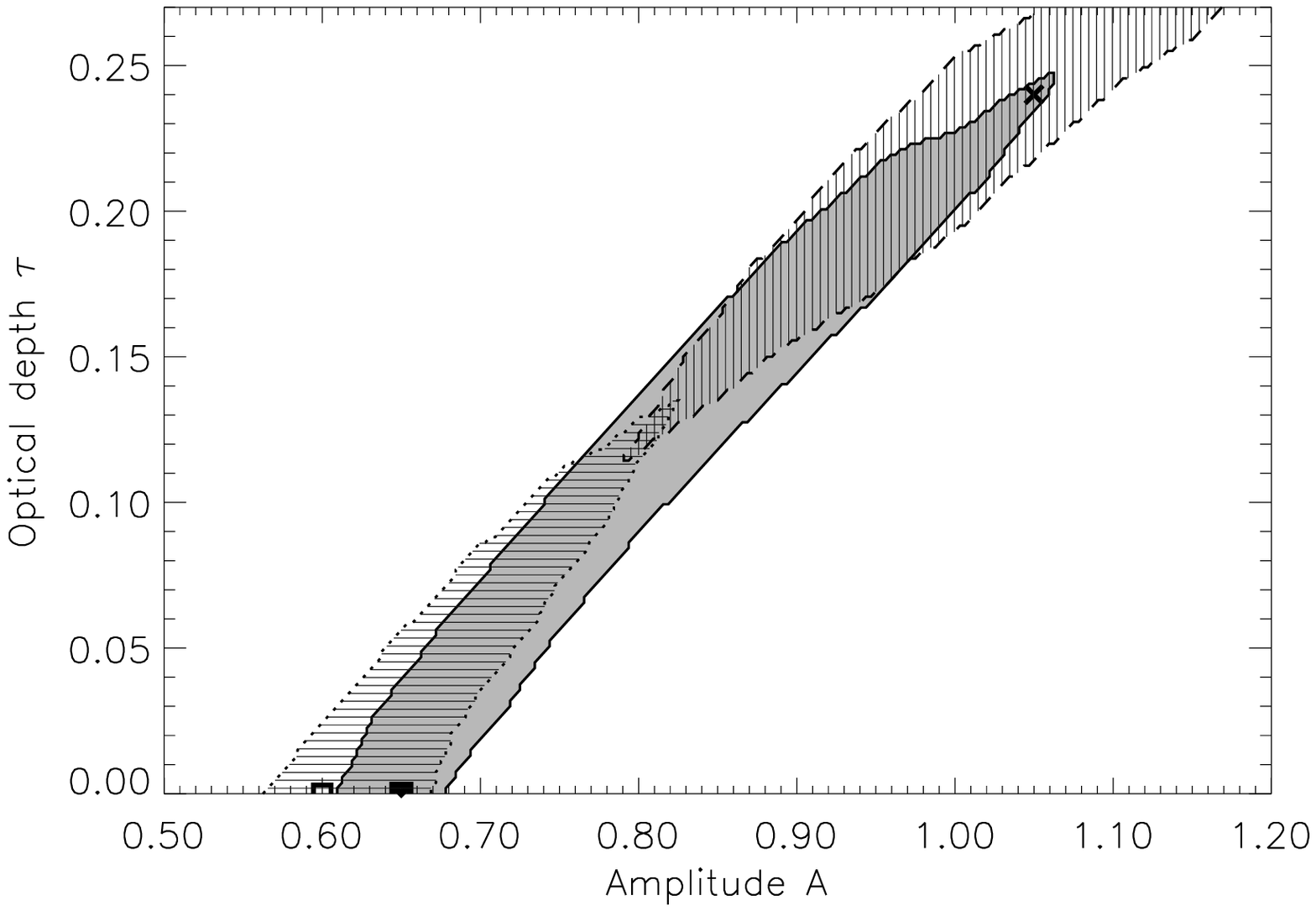,width=7cm,height=7cm}
\psfig {file=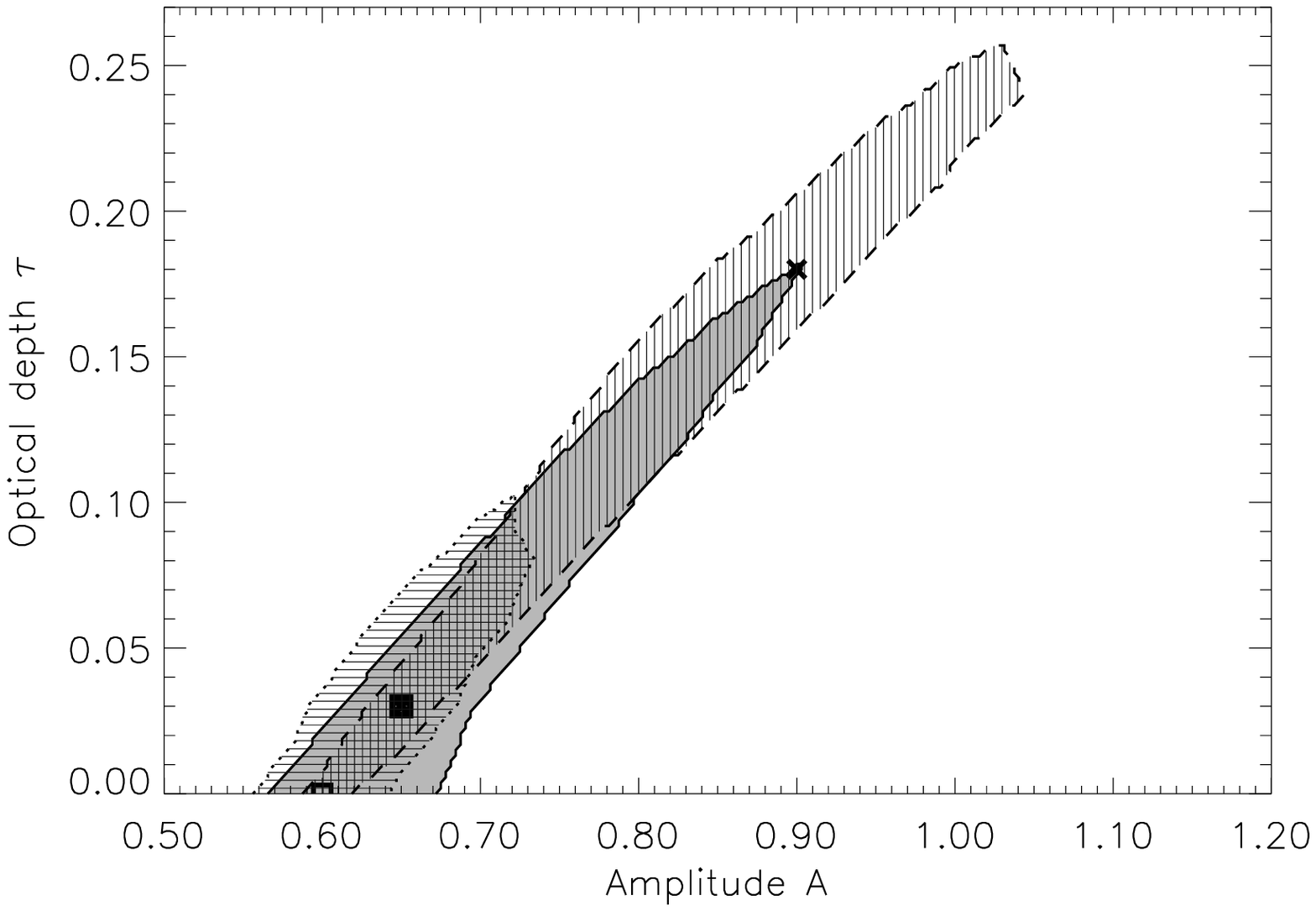,width=7cm,height=7cm}
\caption{The $68\%$ confidence levels for the marginalized likelihood
determined using a Gaussian prior on the spectral index $n$ with $\sigma=0.03$
and centred at either $n=1.0$ (left column) or $n=0.96$ (right
column). The shaded area (solid border line) shows the interval for
the full sky estimate, the area filled with horizontal lines (dotted
border) shows the interval for the northern hemisphere and the area
filled with vertical lines shows the interval for the southern
hemisphere (dashed border). The filled square, open square and cross
show the maximum for the full sky, northern hemisphere and southern
hemisphere respectively. The reference frames are: galactic (upper
row), ecliptic (middle row) and maximum asymmetry (lower row).
Note the increasing disparity between the values derived from the
northern and southern hemispheres as this sequence is descended.
}
\label{fig:plotn}
\end{center}
\end{figure}

\begin{figure}
\begin{center}
\leavevmode
\psfig {file=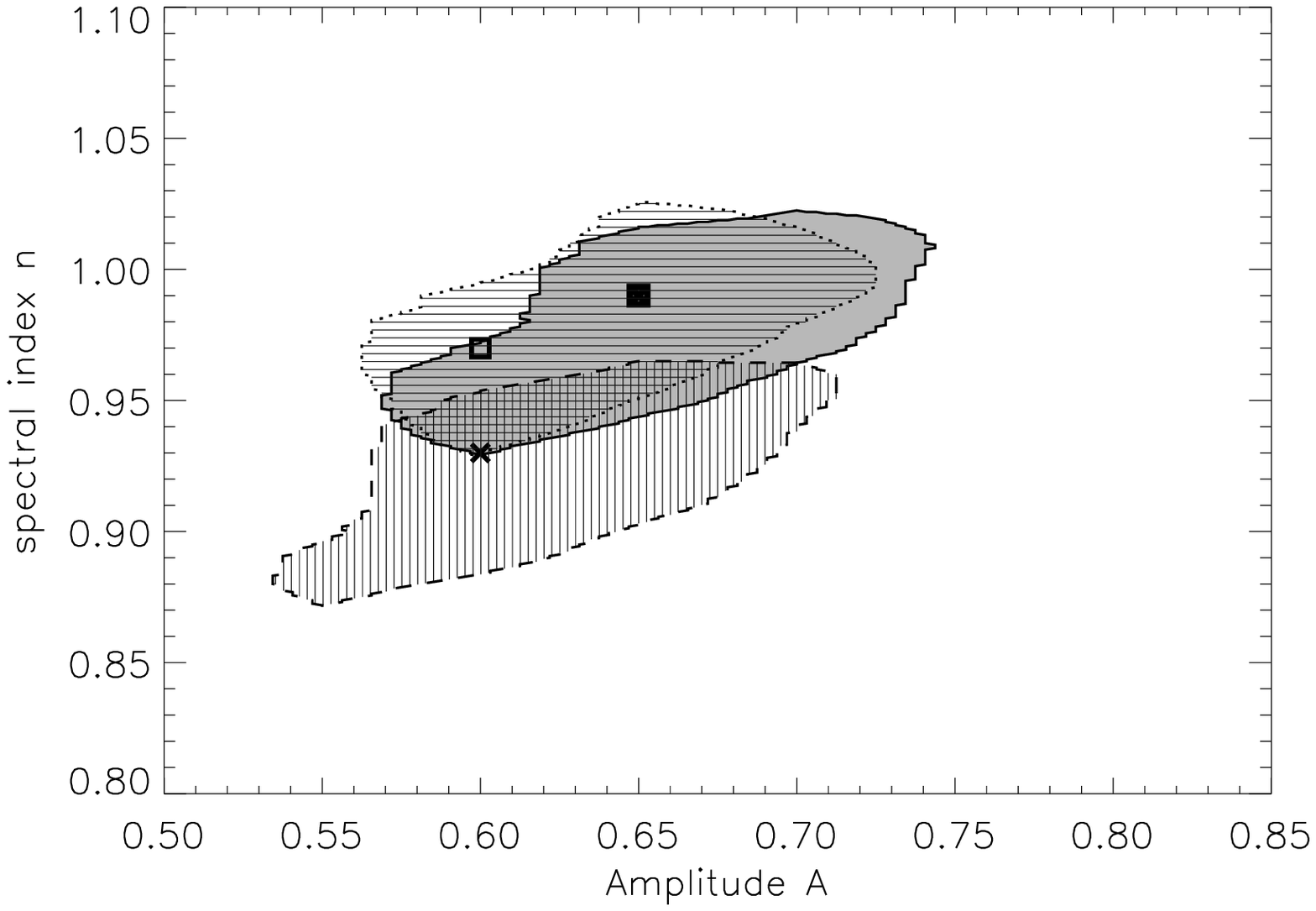,width=7cm,height=7cm}
\psfig {file=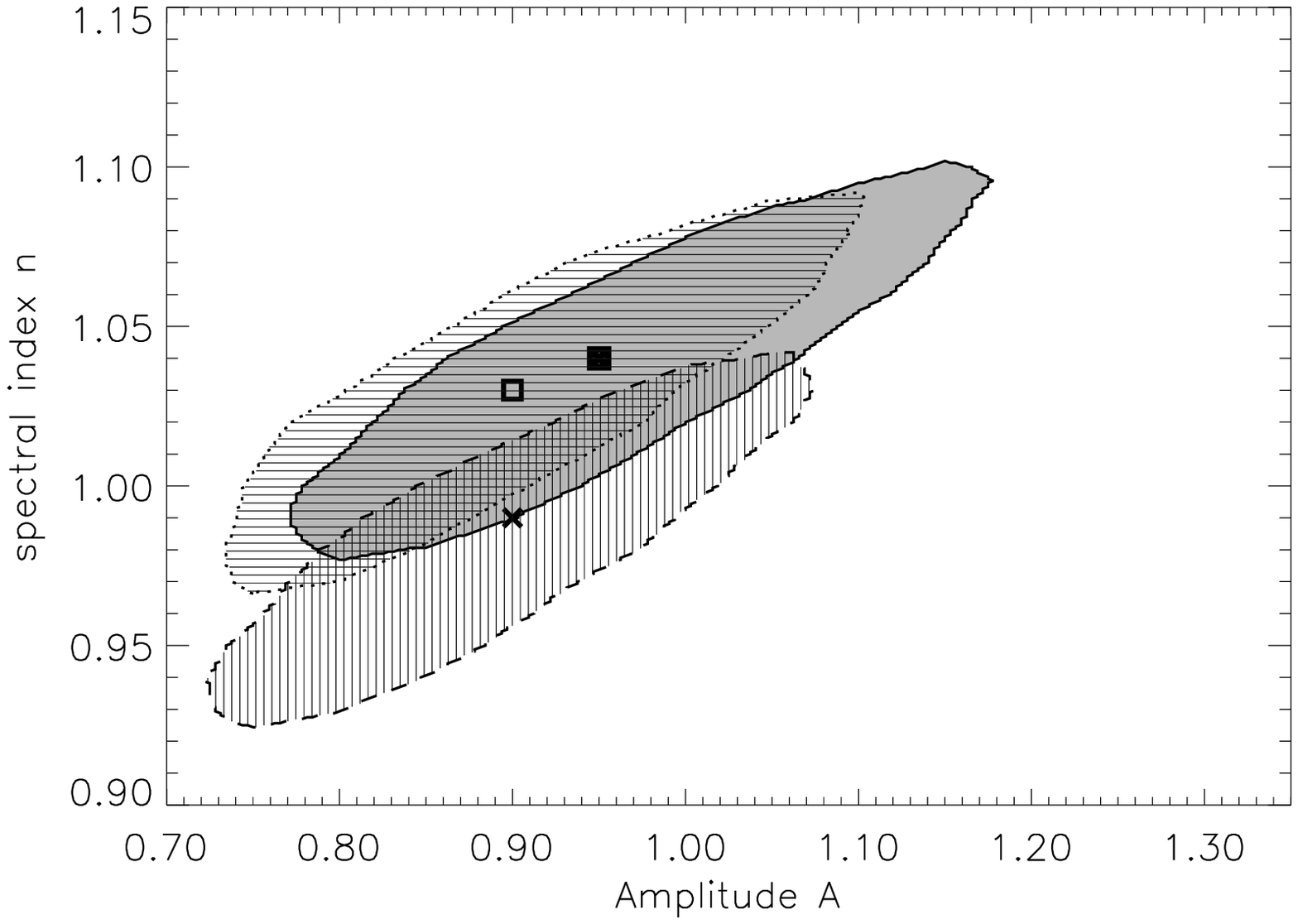,width=7cm,height=7cm}
\psfig {file=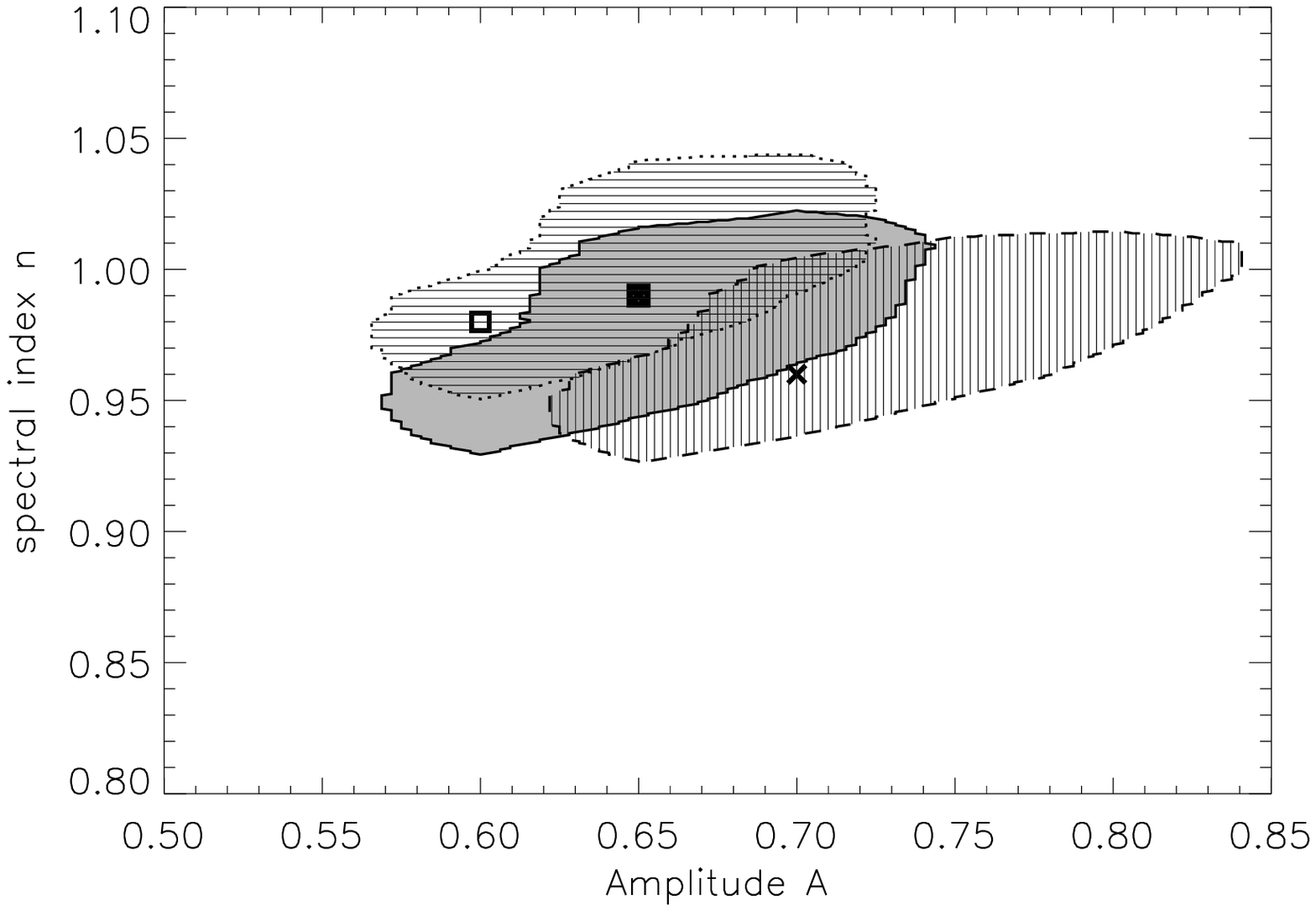,width=7cm,height=7cm}
\psfig {file=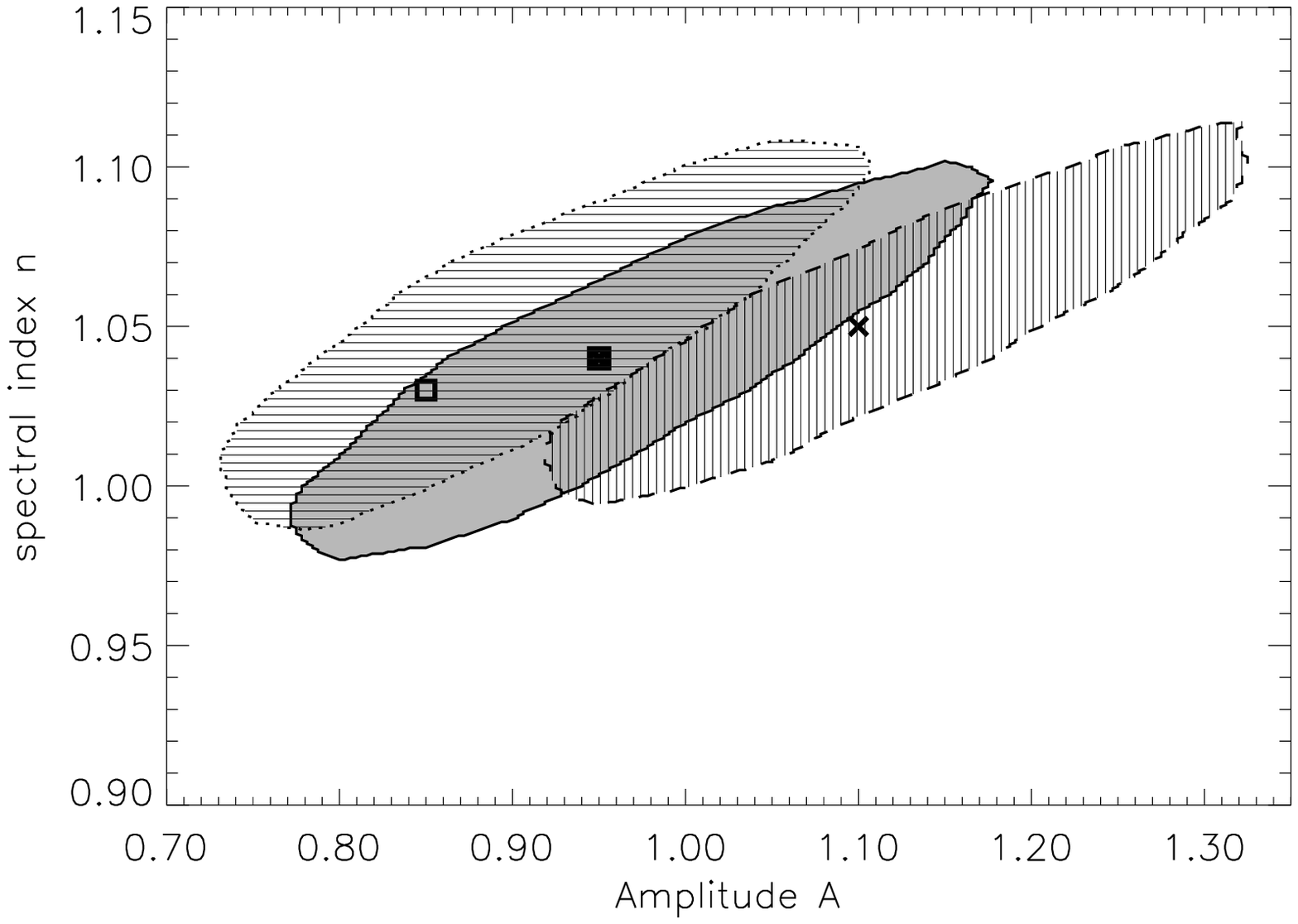,width=7cm,height=7cm}
\psfig {file=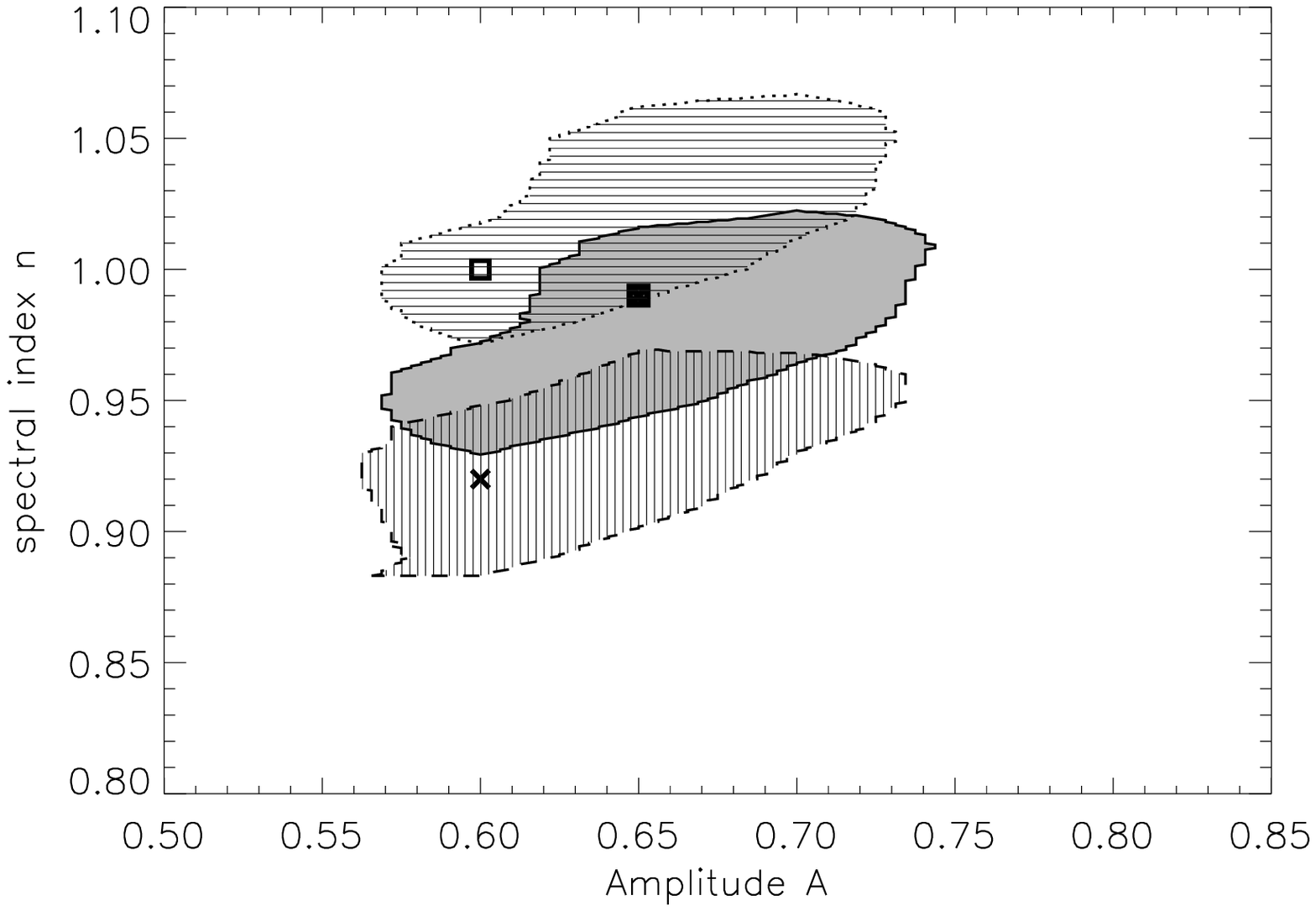,width=7cm,height=7cm}
\psfig {file=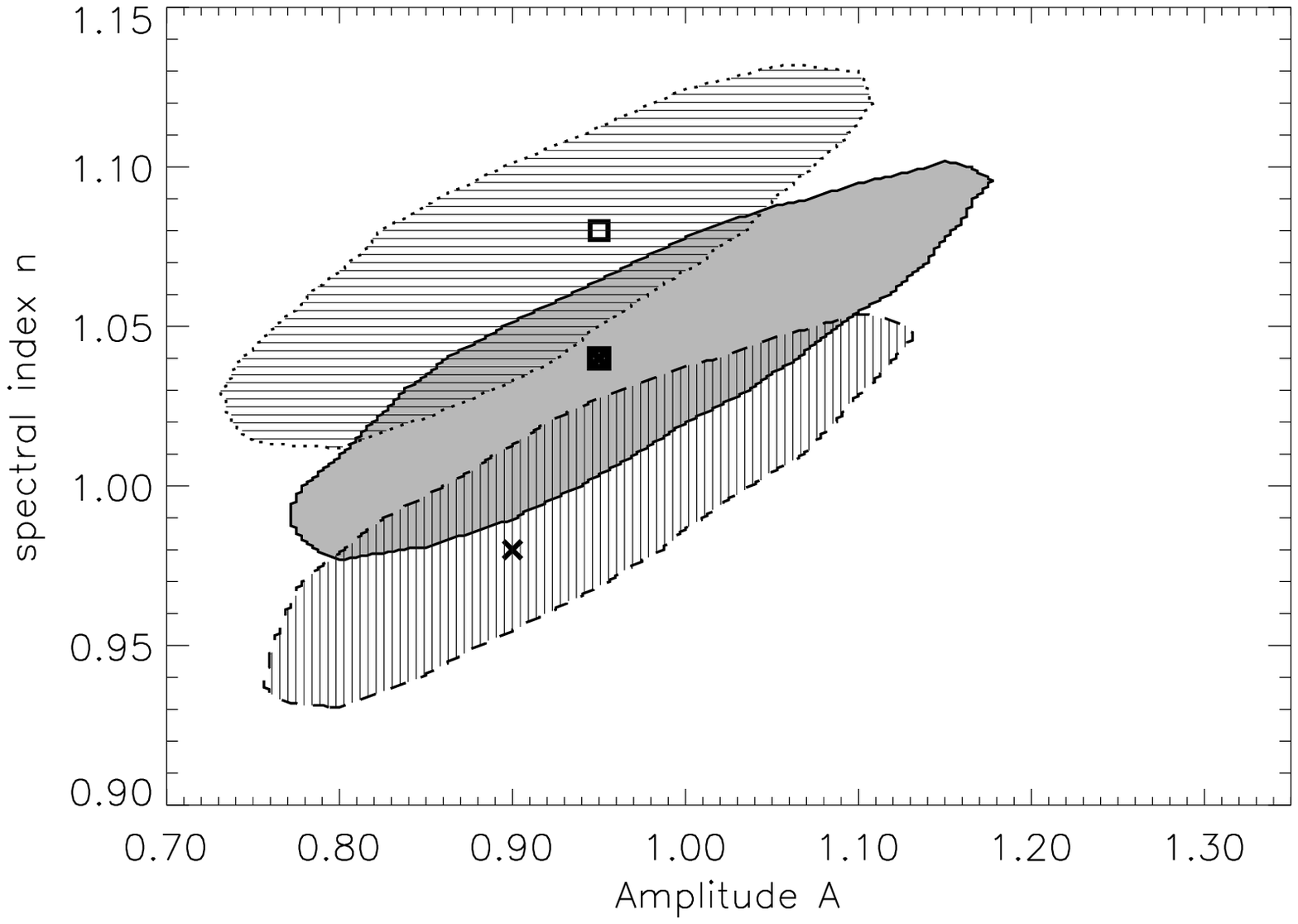,width=7cm,height=7cm}
\caption{The $68\%$ confidence levels for the marginalized likelihood
determined using a Gaussian prior on the optical depth $\tau$ with
$\sigma=0.05$ centred either at $\tau=0.0$ (left column) or
$\tau=0.17$ (right column). The shaded area (solid border line) shows
the interval for the full sky estimate, the area filled with
horizontal lines (dotted border line) shows the interval for the
northern hemisphere and the area filled with vertical lines shows the
interval for the southern hemisphere (dashed border line). The filled
square, open square and cross show the maximum for the full sky,
northern hemisphere and southern hemisphere respectively. The
reference frames are: galactic (upper row), ecliptic (middle row) and
maximum asymmetry (lower row).}
\label{fig:plott}
\end{center}
\end{figure}

For the analysis of the remaining four parameters, it was
found that the adoption of a flat prior resulted in a poorly behaved
likelihood surface, with a large number of false local
maxima. This is a consequence of strong degeneracies between the
parameters. However, by setting stronger priors on two of the parameters, these
degeneracies were reduced and we were able to obtain estimates of the
remaining two parameters. 
For the baryon fraction we adopted
a Gaussian prior with $\Omega_b h^2=0.022$ and
$\sigma=0.001$. This choice was motivated by observations of
the primordial deuterium abundances which give values in the range
$\Omega_b h^2=0.020-0.025$ \cite{d'odorico,o'meara,pettini}.
Alternative Gaussian priors centred at different values in this range
and with different $\sigma$ do not qualitatively affect the results presented below.
Finally, two sets of parameter estimates were computed:
\begin{itemize}
  \item {\bf Gaussian prior on the spectral index $n$}.
    Two different priors were assumed, one centred at $n=1$
    corresponding to a flat Harrison-Zeldovich spectrum, 
    the second at $n=0.96$ as suggested by the best fit \emph{WMAP} 
    spectral index, both with a dispersion $\sigma=0.03$. 
    We found the results robust to changes in
    $\sigma$. $\tau$ and $A$ were then estimated with a flat prior on these two parameters. 
    The results of this procedure as applied to the 6 different hemispheres are shown in table
    (\ref{tab:prin}) and (\ref{tab:prin2}) and figure (\ref{fig:plotn}).
  \item {\bf Gaussian prior on the optical depth $\tau$}. 
    Again two priors were adopted, one centred at $\tau=0$ and one centred at
    the \emph{WMAP} best fit value $\tau=0.17$, both with $\sigma=0.05$. The
    qualitative results and conclusions were found not to change
    significantly under changes in $\sigma$. The results of the parameter
    estimation procedure on the 6 different hemispheres
    are shown in tables (\ref{tab:prit}) and (\ref{tab:prit2}) and figure
    (\ref{fig:plott}).
\end{itemize}

The estimated parameters given in the tables are found in the following way:
\begin{enumerate}
  \item For a given parameter, we found the marginalized likelihood by integrating over all other parameters.
  \item The best fit value quoted in the tables is the value which
    maximises the marginalized likelihood. 
    This value was found to be consistent with the value which  maximises the full 
    likelihood also given in the tables.
  \item The $1\sigma$ and $2\sigma$ ranges given in the tables are those
     for which the integrated marginalized likelihood is $68\%$ 
    and $95\%$ of the total integral respectively. Note that because 
    of degeneracies between parameters, some of these error bars are asymmetric. 
    Forcing a stronger prior (lower $\sigma$) on the other parameters, 
    the error bars become more symmetric.
\end{enumerate}.
The figures were created with following procedure:
\begin{enumerate}
  \item For each possible pair of two parameters we determined the likelihood
    marginalized over the remaining parameters. 
  \item The contours shown in the plot are the areas for which the 
    integral of the two dimensional marginalized likelihood is less than 
    $68\%$ of the total. The shaded zone shows the full sky results, 
    the zones filled with horizontal lines with a dashed border show 
    the results on the northern hemisphere and the zones filled with 
    vertical lines with dotted borders show the results on the southern hemisphere.
  \item The position of the maximum of this two-dimensional likelihood is 
    shown as a filled square (full sky), open square (northern hemisphere) and cross (southern hemisphere).
\end{enumerate}

The most striking feature in these tables is the fact that when
assuming a Harrison-Zeldovich spectrum with a spectral index close
to one, the optical depth in the south is found to be
$\tau\sim 0.23$ whereas the most likely optical depth in the north is consistent with
zero. This holds for all three frames of reference. 
If instead we use a flat prior on $n$ and put a Gaussian
prior on $\tau$, the value of the spectral index becomes significantly
different in the northern and the southern hemispheres with
a larger value in the south, the exact value depending on where the prior on
$\tau$ is centred. Note further that, for the full sky, the estimate
of the optical depth ($\tau=0.16$ assuming a spectral index close to 1)
determined from the marginalized likelihood is close to the
value found by the \emph{WMAP} team, despite the fact that the \emph{WMAP}
results were derived additionally using polarisation data which is more
sensitive to the optical depth than the pure temperature data used
here. It should be noted, however, that as a consequence of strong degeneracies, the
marginalized likelihood in $\tau$ is very flat and the maximum at
$\tau=0.16$ is not a very peaked maximum. This is also seen in figure
\ref{fig:plotn} where the $1\sigma$ contour is very extended in the
$\tau$ direction. This also explains why there is such a large
difference between the estimated value of $\tau=0.16$ using the
marginalized likelihood and the estimate of $\tau=0$ using the full
likelihood as seen in table (\ref{tab:prin}). As will be discussed
below, when excluding the very first multipoles the preferred value of
$\tau$ on the full sky increases.

We will now outline how the results
obtained above can be understood qualitatively 
by considering the manner in which
the parameters of interest, $A$, $\tau$ and
$n$ influence the shape of the spectrum. 
First, we will consider the case when $n$ is fixed and discuss
the variation of $A$ and $\tau$. The optical depth $\tau$ attenuates the
amplitude of the higher multipoles (from $\ell\sim180$ and above) by
a factor of roughly $e^{-2\tau}$. The amplitude and the optical
depth cannot then be estimated separately over this range, 
so an effective resulting amplitude is written as
$Ae^{-2\tau}$. We know that the power spectra
on these scales are consistent between the different
hemispheres, and therefore the estimates of $A$ and $\tau$ should
lie close to the line $Ae^{-2\tau}=constant$ in the $A-\tau$
diagram, as is indeed seen in figure (\ref{fig:plotn}). In order to break
this degeneracy, the shape and amplitude of the lower order multipoles are
needed, since the optical depth has only a small influence on the amplitudes
in this range. Therefore, when fixing the spectrum at
higher multipoles, a greater amplitude for the low multipoles with
respect to the higher ones yields a higher optical depth and
vice versa. In \cite{p3}, it was shown that the amplitude of the
spectrum for the low multipoles ($\ell=2-40$) is higher in the south
than in the north, one could thus expect a higher value of $\tau$ in
the south. This is exactly what we have seen in tables
(\ref{tab:prin}-\ref{tab:prin2}) and in figure (\ref{fig:plotn}).

Now we consider the shape of the spectrum in the first multipoles,
where a strong degeneracy between $\tau$ and $n$ manifests itself.
Specifically, when the optical depth is increased, the shape of the spectrum
is tilted in such a way that the lowest $\ell$ part of the spectrum is
raised with respect to the rest of the low $\ell$ plateau. Lowering
the spectral index $n$ results in similar behaviour. As shown in
\cite {p3}, the southern spectra have this tilt favouring either lower $n$
or higher $\tau$ values. Both of these effects are seen in the
results presented here -- the estimates of $\tau$ with a prior centred at
$n=0.96$ are lower than the corresponding values determined with the prior centred at
$n=1$. Moreover, the estimates of the spectral index $n$ are higher
when the prior on $\tau$ is centred at $\tau=0.17$ than when centred
at $\tau=0$.

In order to understand the significance of the results further,
512 simulations were generated with the effective beam and noise of
the co-added W+V channel, and with the input power spectrum 
specified by the \emph{WMAP} best fit full sky parameters for
a power law index. From these maps, it was determined 
that $5\%$ showed evidence for a difference in $\tau$ between the northern and southern hemispheres
(defined in the maximum asymmetry reference frame of the \emph{WMAP} data) 
as large as that observed here. However, only
one of these 512 maps had a difference in the power spectrum
comparable to that in the \emph{WMAP} data. We then considered
the subset of simulated maps which had,
relative to the \emph{WMAP} results,
a larger difference in $\tau$ but smaller
difference in power spectra between the northern and the southern hemispheres. 
In most cases, the large differences in $\tau$ were due to a high value in one
hemisphere driven by a strong upward fluctuation 
in a few  multipoles.
This is qualitatively different from the \emph{WMAP} behaviour, in which the
spectrum on one of the hemispheres is \emph{systematically} higher over a
large range of $\ell$.

Finally, we note that \cite{p3} found some evidence for possible residual
galactic foreground contamination in the multipole range $\ell=2-4$. 
Therefore, the parameters of interest in this work were re-evaluated
after excluding the multipoles $\ell=2-7$ (this range was appropriate because
of the binning used in this analysis). 
There has been considerable debate in the literature as to 
whether this multipole range is of abnormally low amplitude
relative to the rest of the spectrum (and more particularly relative to the
claimed best fit cosmological model). We do not add to this discussion here,
but recall that such behaviour favours a lower value of
$\tau$ or spectral index $n$, so that on exclusion of
these low-$\ell$ multipoles the preferred value of $\tau$ and $n$
should increase. This is indeed what is observed.
However, the significance of
the asymmetry and the difference in parameters between the north and
the south change only slightly and the results are not reported here.

\begin{center}
\begin{table}
\caption{Estimated parameters with a prior on $n$ centred at $n=1.0$}
\label{tab:prin}
\tiny
\begin{tabular}{|l|c|c|c|c|c|c|c|c|}
Hemisphere & optical depth $\tau$ & $1\sigma$ range ($\tau$) & $2\sigma$ range ($\tau$) & abs. max ($\tau$) & amplitude $A$ & $1\sigma$ range ($A$) & $2\sigma$ range ($A$) & abs. max ($A$) \\
\hline 
Galactic north& 0.01 & 0.0-0.15 & 0.0-0.24 & 0.03 & 0.66 & 0.60-0.85 & 0.57-1.03& 0.65 \\ 
Galactic south& 0.23 & 0.16-0.28 & 0.07-0.31 & 0.21 & 0.97 & 0.85-1.09 & 0.69-1.15 & 0.95 \\
Ecliptic north& 0.0 & 0.0-0.11 & 0.0-0.21 & 0.03 & 0.66 & 0.60-0.78 & 0.57-0.95 & 0.65  \\
Ecliptic south & 0.22 & 0.15-0.28 & 0.06-0.31 & 0.18  & 1.06 & 0.92-1.12 & 0.74-1.27 & 1.00  \\
Max. ass. north& 0.0 & 0.0-0.08 & 0.0-0.18 & 0.03 & 0.65 & 0.59-0.74 & 0.56-0.88 & 0.65  \\
Max. ass. south& 0.24 & 0.17-0.3 & 0.08-0.32 & 0.24 & 1.03 & 0.89-1.12 & 0.71-1.18 & 1.05 \\
Full sky & 0.16 & 0.01-0.22 & 0.0-0.26 & 0.0 & 0.67 & 0.64-0.93 & 0.61-1.10 & 0.65\\
\hline
\end{tabular}
\end{table}
\end{center}

\begin{center}
\begin{table}
\caption{Estimated parameters with a prior on $n$ centred at $n=0.96$}
\label{tab:prin2}
\tiny
\begin{tabular}{|l|c|c|c|c|c|c|c|c|}
Hemisphere & optical depth $\tau$ & $1\sigma$ range ($\tau$) & $2\sigma$ range ($\tau$) & abs. max ($\tau$) & amplitude $A$ & $1\sigma$ range ($A$) & $2\sigma$ range ($A$) & abs. max ($A$) \\
\hline 
Galactic north& 0.0 & 0.0-0.1 & 0.0-0.2 & 0.0 & 0.62 & 0.58-0.73 & 0.56-0.89& 0.60 \\ 
Galactic south& 0.17 & 0.08-0.23 & 0.0-0.26 & 0.15 & 0.82 & 0.68-0.95 & 0.58-1.04 & 0.80 \\
Ecliptic north& 0.0 & 0.0-0.07 & 0.0-0.17 & 0.0 & 0.62 & 0.58-0.71 & 0.56-0.83 & 0.60  \\
Ecliptic south & 0.15 & 0.06-0.22 & 0.0-0.25 & 0.09  & 0.88 & 0.71-1.01 & 0.64-1.15 & 0.80  \\
Max. ass. north& 0.0 & 0.0-0.06 & 0.0-0.14 & 0.0 & 0.61 & 0.58-0.68 & 0.56-0.79 & 0.60  \\
Max. ass. south& 0.18 & 0.09-0.24 & 0.0-0.27 & 0.18 & 0.86 & 0.70-0.98 & 0.59-1.08 & 0.90 \\
Full sky & 0.0 & 0.0-0.12 & 0.0-0.21 & 0.03 & 0.66 & 0.60-0.79 & 0.57-0.96 & 0.65\\
\hline
\end{tabular}
\end{table}
\end{center}

\begin{center}
\begin{table}
\caption{Estimated parameters with a prior on $\tau$ centred at $\tau=0.0$}
\label{tab:prit}
\tiny
\begin{tabular}{|l|c|c|c|c|c|c|c|c|}
Hemisphere & spectral index $n$ & $1\sigma$ range ($n$) & $2\sigma$ range ($n$) & abs. max ($n$) & amplitude $A$ & $1\sigma$ range ($A$) & $2\sigma$ range ($A$) & abs. max ($A$) \\
\hline
Galactic north& 0.98 & 0.95-1.01 & 0.92-1.04 & 0.97 & 0.62 & 0.58-0.68 & 0.55-0.76 &  0.60\\
Galactic south& 0.93 & 0.90-0.96 & 0.87-0.98 & 0.93 & 0.61 & 0.57-0.67 & 0.54-0.76 & 0.60  \\
Ecliptic north& 0.99 & 0.97-1.03 & 0.94-1.06 & 0.98 & 0.64 & 0.59-0.70 & 0.56-0.77 & 0.60   \\
Ecliptic south& 0.96 & 0.95-1.01 & 0.92-1.04 & 0.96 & 0.71 & 0.65-0.79 & 0.62-0.89 & 0.70 \\
Max. ass. north& 1.01 & 0.99-1.05 & 0.96-1.08 & 0.99 & 0.65 & 0.60-0.70 & 0.56-0.77 & 0.60 \\
Max. ass. south& 0.93 & 0.91-0.96 & 0.87-0.99 & 0.93 & 0.62 & 0.58-0.69 & 0.55-0.77 & 0.60  \\
Full sky& 0.96 & 0.95-1.01 & 0.92-1.04 & 0.95 & 0.65 & 0.60-0.71 & 0.57-0.79 & 0.60 \\ 
\hline
\end{tabular}
\end{table}
\end{center}

\begin{center}
\begin{table}
\caption{Estimated parameters with a prior on $\tau$ centred at $\tau=0.17$}
\label{tab:prit2}
\tiny
\begin{tabular}{|l|c|c|c|c|c|c|c|c|}
Hemisphere & spectral index $n$ & $1\sigma$ range ($n$) & $2\sigma$ range ($n$) & abs. max ($n$) & amplitude $A$ & $1\sigma$ range ($A$) & $2\sigma$ range ($A$) & abs. max ($A$) \\
\hline
Galactic north& 1.02 & 0.99-1.07 & 0.95-1.11 & 1.03 & 0.89 & 0.78-1.02 & 0.69-1.15 & 0.90\\
Galactic south& 0.99 & 0.94-1.02 & 0.91-1.06 & 0.99 & 0.88 & 0.77-1.00 & 0.69-1.13 & 0.90  \\
Ecliptic north& 1.06 & 1.01-1.09 & 0.97-1.13 & 1.03 & 0.89 & 0.78-1.02 & 0.69-1.12 & 0.85   \\
Ecliptic south& 1.05 & 1.01-1.08 & 0.97-1.12 & 1.05 & 1.10 & 0.98-1.25 & 0.86-1.36 & 1.10 \\
Max. ass. north& 1.08 & 1.03-1.11 & 0.99-1.15 & 1.08 & 0.90 & 0.79-1.02 & 0.70-1.16 & 0.95 \\
Max. ass. south& 0.98 & 0.95-1.03 & 0.92-1.07 & 0.98 & 0.91 & 0.80-1.04 & 0.72-1.17 & 0.90  \\
Full sky& 1.04 & 1.00-1.08 & 0.96-1.12 & 1.04 & 0.94 & 0.83-1.08 & 0.74-1.21 & 0.95 \\ 
\hline
\end{tabular}
\end{table}
\end{center}

\section{discussion and conclusions}

Previous work has determined the presence of a significant
asymmetric distribution of large scale power in the 
\emph{WMAP} data. Such a result could be suggestive
of incompletely understood foreground residuals
or systematic artifacts, interesting physical effects
taking place at post-recombination, or a more profound
lack of the correct cosmological framework within
which to understand the data. In an attempt to gain
further insight, we have investigated the influence of 
the asymmetry on the inferred values of specific cosmological 
parameters computed in the northern and southern hemispheres of
three frames of reference -- the galactic, the ecliptic and the frame of
reference which maximises the observed asymmetry of the power spectrum. 
It was determined that the parameters which are sensitive to 
variations in the shape of the power spectrum seen in the different
hemispheres are: 
the amplitude of fluctuations $A$, 
the spectral index $n$ 
and the optical depth $\tau$. However, significant degeneracies between
the parameters were present, and in detailed work we elected
to place prior constraints one one parameter, then consider
the behaviour of the remaining two. 

After placing a Gaussian prior on $n$ centred at $n=1$ with
$\sigma=0.03$, we found that the optical depth in the northern
hemisphere was small ($\tau<0.08$ at $1\sigma$ confidence in the maximally
asymmetric frame of reference), whereas in the south the values were of
order $\tau\sim0.23$, inconsistent with $\tau=0$ at the $2\sigma$
level. Therefore, it seems possible that the main contribution to the optical depth
$\tau=0.17$ estimated on the full sky by the \emph{WMAP} team (and independently 
re-estimated here to be $\tau=0.16$) comes from the southern
hemisphere. 
The \emph{WMAP} value for the optical depth also includes information
from the temperature-polarisation, $T\times E$, cross-power
spectrum. A crucial test of our hypothesis is then to make
a similar hemisphere-partitioned analysis of the polarised signal.
A varying optical depth in different directions might occur
if the reionization of the universe were highly
inhomogeneous, but it is not clear how to produce such a scenario. 
An alternate possibility might arise from the presence of a huge gas mass in the southern
hemisphere. This possibility was also discussed by \cite{cruz} in the context
of a strong non-Gaussian cold spot located in this region. However,
the asymmetry in the spectrum remained even when
this spot were masked out. Moreover, to explain such a huge difference
in $\tau$, an enormous amount of gas distributed over a huge area
would be required: assuming that a cluster of galaxies may have an
optical depth of order $\tau\sim0.01$ implies that the gas
required to produce an amplitude of $\tau\sim0.23$ would correspond to 
the content of several clusters along the line of
sight in all directions on the southern hemisphere. If on the other
hand, one assumes that the \emph{WMAP} data are contaminated by some unknown
source, then the surprisingly high value of $\tau=0.23$ might be taken
as an indication that the southern hemisphere is suspect.

Further, the estimate of the amplitude of fluctuations $A$ and the
optical depth $\tau$ using only the temperature power spectrum, are
highly correlated and closely follow the line
$Ae^{-2\tau}=constant$. Consequently, a low estimate of $\tau$ implies
a low estimate of $A$, and vice versa. In the north where the
optical depth is low, the amplitude is also found to be low ($A\sim0.7$)
whereas in the south the amplitude is found to be significantly higher
$A\sim1.0$. As outlined in \cite{spergel}, some weak lensing
experiments and studies using galaxy velocity fields find values of
the amplitude consistent with our northern hemisphere findings. 
This is also the case for a recent estimate of the amplitude using the cluster
mass function obtained from the baryon mass of clusters of galaxies
\cite{vikh}. However, other experiments such as studies of high redshift
clusters and measurements of the SZ effect through the high $\ell$
tail of the CMB power spectrum (again see \cite{spergel} as well as a recent paper by \cite{odman}) 
indicate a higher value for the amplitude, consistent with
our estimate on the southern hemisphere.

Alternatively, if we put a Gaussian prior on $\tau$ and adopt a flat
prior on the spectral index, we find that the spectral index determined in the
northern hemisphere is systematically higher than in the southern
hemisphere, where the exact value depends on where the prior for $\tau$ is
centred. For a low value of $\tau$, we estimate low values
inconsistent with $n=1$ in the southern hemisphere. On the other hand,
if $\tau=0.17$ we find high values inconsistent with $n=1$ in the
northern hemisphere. If we consider that this behaviour is an indication
that one of the hemispheres contains some unknown contamination, one cannot use this result to draw a
conclusion about its location.

Finally, we note that should the asymmetry remain unexplained
in terms of residual or unknown foreground contamination or systematic 
artifacts,
the possibility remains that it may be necessary to invoke
new post-recombination physical processes
or to accept that our current conceptual framework based on the principle
of cosmological isotropy and inflationary-scenario-inspired models
is inadequate. 
We await the release of the \emph{WMAP} 2-year results, and
particularly the polarisation information, to further test
these possibilities.

\section*{Acknowledgements}

We are grateful to D. Marinucci, P. Mazzotta, A. Vikhlinin and
S.D.M. White for useful discussions and comments. FKH acknowledges
financial support from the CMBNET Research Training Network. We
acknowledge use of the HEALPix \cite{healpix} software and analysis
package and the CMBfast software package \cite{cmbfast} for deriving
the results in this paper. This research used resources of the
National Energy Research Scientific Computing Center, which is
supported by the Office of Science of the U.S. Department of Energy
under Contract No. DE-AC03-76SF00098. We acknowledge the use of the
Legacy Archive for Microwave Background Data Analysis
(LAMBDA). Support for LAMBDA is provided by the NASA Office of Space
Science.

\end{document}